\documentclass[12pt, reqno]{amsart}

\usepackage[body={6.3in,9.3in}]{geometry}
\usepackage{verbatim}
\usepackage[ansinew]{inputenc}

\usepackage{textcomp}
\usepackage{latexsym}
\usepackage[usenames]{color}
\usepackage{amsmath,amsfonts,amssymb,amsthm}
\usepackage{graphicx}
\usepackage{longtable}
\usepackage{bbm}

\usepackage{setspace}
\usepackage{array}

\def\qed{\hfill{\raggedleft{\hbox{$\Box$}}} \smallskip}

\def\R{\mathbb{R}}
\def\K{\mathcal{K}}

\def\I{\mathbf{I}}

\def\ST{\mathcal{ST}}

\newcommand{\ds}{\displaystyle}

\theoremstyle{plain} \newtheorem{lem}{Lemma}
\theoremstyle{plain} \newtheorem{prop}[lem]{Proposition}
\theoremstyle{plain} \newtheorem{thm}[lem]{Theorem}
\theoremstyle{plain} \newtheorem{cor}[lem]{Corollary}
\theoremstyle{plain} 
\theoremstyle{definition} 
\theoremstyle{definition}
\theoremstyle{definition} 
\theoremstyle{definition} 
\theoremstyle{definition}

\newlength\savedwidth

\def\1{\mathbf{1}}

\author{Ngoc Mai Tran}
\address{Department of Statistics, UC Berkeley, CA 94720, USA}
\email{tran@stat.berkeley.edu}
\thanks{The author would like to thank Lek-Heng Lim, Noureddine El Karoui and Bernd Sturmfels for helpful comments and advices. Research on this project was supported by the National Science Foundation (DMS-1057064) and (DMS-0968882)}

\title{HodgeRank is the limit of Perron Rank}

\begin{document}
\begin{abstract}{We study the map which takes an elementwise positive matrix to the k-th root of the principal eigenvector of its k-th Hadamard power. We show that as $k$ tends to 0 one recovers the row geometric mean vector and discuss the geometric significance of this convergence.  In the context of pairwise comparison ranking, our result states that HodgeRank is the limit of Perron Rank, thereby providing a novel mathematical link between two important pairwise ranking methods.}
\end{abstract}
\maketitle
Let $X = [X_{ij}]$ be a $n \times n$ matrix, $\K := \R_+^{n \times n}$ be the open cone of elementwise positive matrices. For $X \in \K$, let $v(X), \lambda(X)$ denote its unique principal eigenvector-eigenvalue pair in $\mathbb{PR}^{n-1} \times \R$, $X^{(k)} := [X_{ij}^k]$ be its $k^{th}$ Hadamard power. It is known that if the tropical max-times eigenvector $m(X)$ of $X$ is unique in $\mathbb{PR}^{n-1}$, then it is the coordinate-wise limit of the sequence $v(X^{(k)})^{1/k}$ as $k \to \infty$ \cite{gaubert}.
$$\lim_{k \to \infty}v(X^{(k)})^{1/k} = m(X) \label{eqn:trop.conv}$$
Our first theorem states that the same sequence converges coordinate-wise to the row geometric mean of $X$ as $k \to 0$. 
\begin{thm}\label{thm:conv}
For $X \in \K$, define $h(X)_i = (\ds\prod_{j=1}^nX_{ij})^{1/n}$. Then
$$\lim_{k \to 0}v(X^{(k)})^{1/k} = h(X) \label{eqn:h.conv}$$
\end{thm}

This result plays an important role in the context of pairwise ranking, where $X$ is restricted to the subvariety of \emph{multiplicative comparison} or \emph{symmetrically reciprocal} matrices $\{X \in \K: X_{ij} = 1/X_{ji}\}$ \cite{dahl, elsner10, farkas, fuzzy}. Here $X_{ij}$ measures the multiplicative preference of $i$ over $j$. A ranking algorithm takes $X$ as input and returns a score vector $s \in \mathbb{PR}^{n-1}$ by which the items are ranked. The triple $v(X), m(X), h(X)$ are outputs of three ranking algorithms: Perron Rank \cite{fuzzy, perron}, Tropical Rank \cite{elsner, elsner10} and HodgeRank \cite{crawford, lekheng}. Perron Rank plays a fundamental role behind the Analytic Hierarchical Process \cite{fuzzy}, a ranking procedure extensively applied in decision making. On the other hand, HodgeRank is closely related to many pairwise ranking algorithms in the rank learning literature of computer science \cite{meier, meierLearnLabels, lekheng}. A number of papers have been devoted to their comparisons \cite{dong, elsner, farkas, lekheng, saari, fuzzy, nmt}. However, Theorem \ref{thm:conv} is the first to show that HodgeRank can be obtained as the limit of Perron Rank, thus providing a mathematical tool for comparing pairwise ranking methods in computer science to those often used in decision making.

Hadamard powers of a multiplicative comparison matrix arise naturally when one attempts to apply Perron Rank to \emph{additive comparison} matrices, those which arise often in computer science applications \cite{meier, hochbaum, meierLearnLabels, lekheng}. These are skew-symmetric matrices $\wedge_2\R^n$ where $A_{ij}$ measures the additive preference of $i$ over $j$. They are in natural bijection with multiplicative comparison matrices via the exponential/log map, however, the bijection is not unique since one is free to choose the base value. Changing the exponential base is equivalent to multiplying the additive matrix $A$ by a positive constant, which can be interpreted as changing the measurement unit. This corresponds to taking Hadamard power of the mupliticative matrix $X$. Thus, if the input is an additive matrix $A$ and \emph{a priori} no base is preferred, one should consider all the Hadamard powers $[\exp(kA_{ij})]$ when applying Perron Rank $v: X \mapsto v(X)$. This yields the \emph{Perron family} of ranking methods $V_k: X \mapsto v(X^{(k)})^{1/k}$ for any $k \in (0, \infty)$, with HodgeRank and Tropical Rank appear as limiting cases. We shall use the notation $\tilde{V}_k: A \mapsto 1/k\log v([\exp(ka_{ij})])$ to denote its log version. 

It can easily be checked that the Perron family, HodgeRank and Tropical Rank are projections onto the set of \emph{strongly transitive} matrices \cite{saari} or \emph{consistent} \cite{fuzzy} matrices, which are rank one matrices identified with the score vectors in $\mathbb{PR}^{n-1}$ via $w \leftrightarrow [w_i/w_j]$. Its image in $\wedge_2\R^n$ under the log map is the subspace spanned by the matrices $\sum_{j=1}^ne_{ij} - \sum_{k=1}^ne_{ki}$ for $i = 1, 2, \ldots, n$, where $e_{ij}$ is the matrix with 1 in the $(i,j)$-th entry and 0 else, denoted $\mathcal{ST}$. These matrices are so named since for all triples $i, j, k$ the relation $X_{ij} = X_{ik} \cdot X_{kj}$ holds, which is natural if one assumes that the comparison matrix is obtained from the score in the error-free. Thus one central question in pairwise ranking is the \emph{score recovery problem}: suppose there is a true score $w$ and one observes $X$, a perturbed version of the true comparison matrix $[w_i/w_j]$. Which method is `best', in some sense, at recovering $w$? By specifying an objective function, such as number of items ranked correctly, one can convert the comparison of methods into an optimization problem. This formulation is commonly found in the broader area of rank learning \cite{DMJ, meier, hochbaum}, a rich litearture on the study of algorithms for ranking with emphasis on large datasets, predicting the rank of new items and ranking the top items correctly(see, for example, \cite{meier} and references therein).

A large class of these optimization problems can be rephrased in $\wedge_2\R^n$ as geometric questions on the images under the projection $\tilde{V}_k$ of level sets of the noise distribution. 
The second result of our paper, Theorem \ref{thm:ker}, describes the fibers of this projection. It also contributes towards solving whether there exists an objective function in which Perron Rank for $k \neq 0$ or $\infty$ is the optimal solution over all possible ranking algorithms. This question arised in the literature since HodgeRank is known to be the $\ell_2$-minimizer and Tropical Rank is known to be a special point on the set of $\ell_\infty$-minimizers to $\mathcal{ST}$ \cite{elsner10, lekheng}. 

\begin{thm}\label{thm:ker}
 Let $\1 = (1,\ldots,1)^T$ be the all one vector, $\mathbf{0} = (0, \ldots, 0)^T$ the all-zero vector.
 For all $s \in \R / (1, \ldots, 1)$, as a set, the fibers of the map $\tilde{V}_k$ satisfy
$$\tilde{V}_k^{-1}(s) = [s_i - s_j] + \tilde{V}_k^{-1}(\mathbf{0}).$$ 
The zero-fiber $\tilde{V}_k^{-1}(\mathbf{0})$ can be decomposed as
\[ \tilde{V}_k^{-1}(\mathbf{0}) = \bigoplus_{i=1}^n S_i(k) + \R\cdot \1\1^T,  \]
where the $S_i(k)$ for $i = 1, \ldots, n$ are orthogonals, with
\begin{align*}
S_i(k) & := \{a_{i\cdot} \in \R^n: a_{ij} < 0, \sum_{j=1}^n\exp(ka_{ij}) = 1\} \mbox { for } k \in (0, \infty), \\
S_i(\infty) &:= \{a_{i\cdot} \in \R^n: a_{ij} \leq 0, a_{ij} = 0 \mbox{ for at least one } j = 1, \ldots, n\}, \\
S_i(0) &:= \{a_{i\cdot} \in \R^n: \sum_{j=1}^na_{ij} = 0\} 
\end{align*}
\end{thm}
Since the fibers $\tilde{V}_k^{-1}(s)$ are just translations of the zero-fiber by $[s_i - s_j]$, Theorem \ref{thm:ker} implies that the optimal $k^\ast$ in the score recovery problem considered is independent of the true additive score $s$. For large $k$, the zero-fiber is the real amoeba of the tropical (max-plus) eigenvector map \cite{bernd.trop}. For $k = 0$, the zero-fiber is the set of matrices decomposing into the all-one matrix plus a matrix whose column sum is zero, which is the zero-fiber of the HodgeRank map. Thus we obtain a crude version of Theorem \ref{thm:conv}.  

In summary, Theorem \ref{thm:conv} and its geometric cousin, Theorem \ref{thm:ker} are new mathematical results linking HodgeRank and Perron Rank, two popular methods used in pairwise ranking. This converts a large class of the score recovery problem into a parametric optimization question with geometric interpretations, contributing towards the solving of specific instances of this problem.  

\section{Proofs}\label{sec:proof.conv}
\subsection{Proof of Theorem \ref{thm:conv}}
Equip $\mathbb{R}^n$ with the Euclidean norm. Let $\|\cdot \|$ denote the operator norm if the argument is a matrix, and the $\ell_2$ norm if the argument is a vector. Let $\I$ be the $n \times n$ identity matrix. Our proof of Theorem \ref{thm:conv} relies on the following proposition, which gives a linear approximation to the `error term' of the principal eigenvector when $X$  is not far from being a strongly transitive matrix. 
\begin{prop}\label{prop:zwei}
Fix a vector $s \in \R_+^p$. Let $\xi := [\xi_{ij}] := [s_j/s_i \cdot X_{ij}]$. Suppose there exists a constant $\kappa \geq 1$ such that the $n \times n$ matrix $\Xi = \xi - \kappa\1\1^T - (1-\kappa)\I$ 
satisfies $\rho := (\frac{2\|\Xi\|}{n\kappa - 2\|\Xi\|})^2 < \frac{1}{2}.$ Then
\begin{equation} 
\ds v(X) = s \cdot (\1 + \frac{1}{\kappa n}(r - \bar{r}\1) + \epsilon) \label{eqn:v}
\end{equation}
where $r = (\Xi \cdot \1)$ is the row sum of $\Xi$, $\bar{r} = \frac{1}{n}\sum_{j=1}^nr_j$ is its mean, and 
$\|\epsilon\|~< ~\frac{\rho}{1-\rho}\cdot~ \frac{\|\Xi\|}{\kappa\sqrt{n}}$. 
\end{prop}
This proposition is interesting in itself. One could think of $s$ as the true score vector, $[s_i/s_j]$ as the true multiplicative comparison matrix, and $\xi$ as the multiplicative perturbation with centered version $\Xi$, so defined since $v(\kappa\1\1^T + (1-\kappa)\I) = \1$ for all $\kappa \geq 1$. If $\Xi \equiv 0$ then the perturbation does not affect the principal eigenvector, and $v(X)$ is exactly $s$. The proposition states that if the centered perturbation $\Xi$ is small, as measured by $\rho$ and $\frac{\|\Xi\|}{\kappa\sqrt{n}}$, then $v(X)$ differs from the true multiplicative score $s$ only by a linear factor plus a lower order term. Unfortunately this proposition cannot be applied easily in practice since it requires knowledge of $\kappa$, which depends on the choice of $s$ and in general neither are easy to find. For the proof of Theorem \ref{thm:conv} it is sufficient to choose $\kappa = 1$. 
\vskip12pt
\emph{Proof of Proposition \ref{prop:zwei}.}
By \cite[Lemma 2.2]{nmt}, $v(X) = s\cdot v(\xi)$, hence one can assume without loss of generality that $s \equiv \1$. Write 
$\xi = \kappa \1\1^T + (\Xi + (1-\kappa)\I)$. Then $\kappa \1\1^T$ is a rank one matrix with one non-zero eigenvalue $p\kappa$, corresponding to the normalized eigenvector $\frac{1}{\sqrt{n}}\1$. Let $Y \in \R^{n \times (n-1)}$ be an orthonormal basis of the zero eigenspace of $\kappa \1\1^T$. From standard results in perturbed linear operators (see, for example, \cite[Theorem 2.7]{plo.book}), $\rho < \frac{1}{2}$ implies $v(\xi) = \frac{1}{\sqrt{n}}\1 + YP + \frac{\epsilon}{\sqrt{n}},$
where $P := \frac{1}{n\sqrt{n}\kappa}Y^T(\Xi + (1-\kappa)\I)\1$, and the error term is bounded by 
\[ \|\epsilon\| \leq \frac{\rho}{1-\rho}\cdot \frac{\|\Xi\|}{n\kappa}.\] 
Since $YY^T = \I - \frac{1}{n}\1\1^T$, and since $v(\xi)$ is only defined up a multiplicative constant, we have
\begin{align*}
v(\xi) &= \frac{1}{\sqrt{n}}\left( \1 + \frac{1}{n\kappa}(\textbf{I} - \frac{1}{n}\1\1^T)(\Xi+(1-\kappa)\I)\1 +\epsilon \right) \\
&= \1 + \frac{1}{n \kappa }(\Xi\1 - \frac{1}{n}\1\1^T\Xi\1) + \epsilon = \1 + \frac{1}{n\kappa}(r - \bar{r}) + \epsilon
\end{align*}
\qed
\vskip12pt 
\noindent
\emph{Proof of Theorem \ref{thm:conv}.}
Under the notation of Proposition \ref{prop:zwei}, let $\kappa = 1$, $s \equiv \1$, so $X = \xi$. Define $A = [A_{ij}] = [\log X_{ij}]$. Then
$$ \Xi^{(k)}_{ij} = \exp(kA_{ij}) - 1 = k\cdot\left(A_{ij} + \sum_{t \geq 2}\frac{k^{t-1}A_{ij}^t}{t!}\right). $$
Hence for $k$ close to $0$, $\|\Xi^k\| = k\|A\| + o(k)$ and $\frac{\rho}{1-\rho} = O(k^2)$, and Proposition \ref{prop:zwei} applies. Note that $r = \Xi^{(k)} \1 = k\cdot A\1 + O(k^2)$, $\bar{r}\1 = k \cdot \1\1^TA\1 + O(k^2)$, and $\epsilon = O(k^3)$. After applying the logarithm to Equation \ref{eqn:v}, Taylor expansion gives
$$
\lim_{k\to 0}\frac{1}{k}\log v(X^k) = \lim_{k\to 0}\left(\frac{1}{kn}r - \frac{1}{kn}\bar{r} + O(k^2)\right) = \frac{1}{n}A\1 - \frac{1}{n}\1\1^TA\1 = \frac{1}{n}\log h(X)
$$
since the HodgeRank vector is defined only up to additive constants. \qed
\subsection{Proof of Theorem \ref{thm:ker}}\label{sec:proof.ker}
It is sufficient to prove equivalent statements on $\K$ for the map $V_k$. The fibers of this map are the set of positive matrices whose principal eigenvector is a fixed vector $w \in \R_+^n$, that is
\[ \K(w) := \ds\bigsqcup_{\mu \in (0,\infty)}\K(w, \mu) := \bigsqcup_{\mu \in (0,\infty)} \{X \in \K: v(X) = w, \lambda(X) = \mu\} \]
where $\sqcup$ denote disjoint set union. We call this the \emph{one-dimensional real positive Kalman variety}, motivated by the definition of the Kalman variety in Ottaviani and Sturmfels \cite{bernd}. In general real Kalman varities are difficult to characterize, however the one-dimensional positive variety admits a simple description.
\begin{cor}\label{kalman}
For each fixed pair $(w, \lambda)$, let $\Psi_{w, \lambda}: \K \to \K$ be the map $\ds [X_{ij}] \mapsto [Y_{ij}] :=  [\frac{X_{ij}w_j}{\lambda w_i}]$. Then 
$$\Psi_{w,\lambda}(\K(w, \lambda)) = \bigoplus_{i=1}^n(\Delta_{n-1})_i$$ 
where each $(\Delta_{n-1})_i$ is the interior of the $(n-1)$-dimensional simplex on the $n\cdot(i-1)+1$ to $n\cdot i$ coordinate of $\R^{n \times n}$. 
\end{cor}
\noindent
\emph{Proof of Theorem \ref{thm:ker}.}
The first statement follows by \cite[Lemma 2.2]{nmt} (which is a direct computation). By Corollary \ref{kalman}, $V_k^{-1}(\1) = \mathbb{R} \cdot \bigoplus_{i=1}^n(\Delta_{n-1})_i$. By taking $\log$, one obtains the stated result for $\tilde{V}_k^{-1}(\mathbf{0})$. The case for $k = \infty$ follows from the convergence of the real amoebas to the zero set of the tropical eigenvectors map. As $k$ approaches 0, each component $S_i(k)$ is flattened, and a little calculation shows that the each component converges, up to a translation by a large constant times the $(1, \ldots, 1)$ vector, to its tangent plane at $a_{i1} = \ldots = a_{in}$, which has equation $\sum_{j=1}^na_{ij} = 0$. This proves the theorem. \qed

\bibliographystyle{plain}
\bibliography{references}
\end{document}